\documentstyle[prd,aps,preprint]{revtex} 

\def\greaterthansquiggle{\raise.3ex\hbox{$>$\kern-.75em\lower1ex\hbox{$\sim$}}}
\def\lessthansquiggle{\raise.3ex\hbox{$<$\kern-.75em\lower1ex\hbox{$\sim$}}}
\newcommand{\bdi}{\begin{displaymath}}
\newcommand{\edi}{\end{displaymath}}
\newcommand{\bfi}{\begin{figure}}
\newcommand{\efi}{\end{figure}}

\newcommand{\beq}{\begin{equation}}
\newcommand{\eeq}{\end{equation}}

\newcommand{\gaM}{\gamma^{\mu}}

\newcommand{\gaf}{\gamma_{5}}

\newcommand{\tr}{\mbox{tr}}

\newcommand{\beqa}{\begin{eqnarray}}
\newcommand{\eeqa}{\end{eqnarray}}
\newcommand{\no}{\nonumber}

\newcommand{\ra}{\rightarrow}

\newcommand{\wt}{\widetilde}

\newcommand{\dsla}{\partial\hspace{-6pt} /  }  
\newcommand{\Asla}{A\hspace{-6.5pt}  /  }

\begin{document}

\setcounter{page}{0}
\def\footnoterule{\kern-3pt \hrule width\hsize \kern3pt}
\tighten
\title{CONSISTENT AND COVARIANT COMMUTATOR ANOMALIES IN THE CHIRAL
SCHWINGER MODEL\thanks
{This work is supported by a Schr\"odinger Stipendium of the Austrian FWF and
in part by funds provided by the U.S.
Department of Energy (D.O.E.) under cooperative 
research agreement \#DF-FC02-94ER40818.}}

\author{Christoph Adam\footnote{Email address: {\tt adam@ctp.mit.edu, 
adam@pap.univie.ac.at}}}

\address{Center for Theoretical Physics \\
Laboratory for Nuclear Science \\
and Department of Physics \\
Massachusetts Institute of Technology \\
Cambridge, Massachusetts 02139 \\
and \\
Inst. f. theoret. Physik d. Uni Wien \\
Boltzmanngasse 5, 1090 Wien, Austria \\
{~}}

\date{MIT-CTP-2662,~ {~~~~~} September 1997}
\maketitle
\thispagestyle{empty}

\begin{abstract} 
We derive all covariant and consistent divergence and commutator anomalies of
chiral QED$_2$ within the framework of canonical quantization of the fermions.
Further, we compute the time evolution of all occurring operators and
find that all commutators evolve canonically. We comment on the relation of 
our results to the finding of a nontrivial $U(1)$-curvature in gauge-field
space.

\medskip


\end{abstract}
\vspace*{\fill}

\pacs{}

\section{Introduction}

Chiral gauge theories are anomalous when the fermions are quantized. These
anomalies have several, wellknown consequences. The divergence of the gauge 
current deviates from its canonical value by a certain polynomial in the
gauge field (``anomaly'') \cite{Ad1}-\cite{Bertl1}. 
Further, the commutators of charge densities \cite{JJ1} and of the generators
of time-independent gauge transformations (the ``Gauss law operators'')
acquire anomalous contributions, too \cite{Fad1}-\cite{Jack1}. 
In addition, there exist two versions, 
namely the consistent and covariant ones, of both the divergence and 
Gauss law commutator anomalies \cite{BZ1,Bertl1}. 
All these anomalies are determined by geometrical or cohomological
considerations \cite{Sto1}-\cite{Tsu1}. 

However, whereas the complete Gauss law commutator is fixed
in this way, the situation is less clear for the individual components of
the Gauss law operator $G$,
\beq
G(x^1)=\delta (x^1) -iJ^0_+ (x^1)\, ,\quad \delta (x^1)=\partial_{x^1}
\frac{\delta}{e\delta A_1 (x^1)}
\eeq
where $\delta (x^1)$ generates gauge transformations on the gauge fields,
and $J^0_+ (x^1)$, the zero component of the chiral gauge current, acts
on fermions (we have already restricted to our two-dimensional, abelian
model in (1)).
The results for the commutators of the individual components of $G$ depend
both on the computational scheme and on whether one is dealing with
the consistent or covariant case. Further, in most computations VEVs
are computed instead of the operator relations themselves (e.g. by the
use of the BJL limit, or by covariantly regularized gauge current VEVs, or
by a generalized point splitting method \cite{Jo1}-\cite{DT2}). 

In this paper we shall
follow a different approach, by using a method that was introduced in
\cite{IST1,IST2} for the computation of the covariant anomaly. We shall 
construct the second quantized fermionic operators in an external gauge
field both in the interaction and Heisenberg pictures. This will 
enable us to compute the consistent and covariant divergence anomalies
and all the commutators of the Gauss law operator components.
We shall find that all commutators evolve canonically under time
evolution. Finally, we shall comment on the relation of our results to
the findings of a nontrivial $U(1)$-connection and curvature for the
functional derivative $(\delta / \delta A_1 (x^1))$ acting on fermionic
Fock states \cite{NS1}-\cite{Sem1}.

At first sight, it might seem to be a strange idea to discuss consistent
and covariant anomalies for the simple model of chiral QED$_2$, where
both the consistent and covariant anomalies are gauge invariant expressions,
but, nevertheless, there is a difference \cite{IST1,IST2}. This may be
easily inferred, e.g., from the effective action $W[A]$ of the chiral
Schwinger model \cite{Jack1,JR1,AAR},
\beqa
W[A] &=& \frac{ie^2}{4\pi}\int d^2 xd^2 y A_\mu (x)\frac{1}{\Box}(x-y)
[\partial^\mu \partial^\nu -\Box g^{\mu\nu} \no \\
&& -\frac{1}{2}(\epsilon^{\nu\alpha}\partial^\mu \partial_\alpha +
\epsilon^{\mu\alpha}\partial^\nu \partial_\alpha )]A_\nu (y)
+\int d^2 xA_\mu (x) C^{\mu\nu}A_\nu (x).
\eeqa
Here $C^{\mu\nu}$ accounts for the possibility of adding local counterterms 
to the effective action, and we choose $C^{\mu\nu}$ symmetric, because
only the symmetric part of $C^{\mu\nu}$ will contribute to VEVs upon
functional differentiation w.r.t. $A_\mu$. We shall not assume $C^{\mu\nu}$
to be Lorentz covariant in general, as we want to relate to the canonical
formalism where manifest Lorentz covariance is absent. Setting $C^{\mu\nu}
=0$ for the moment we find for the VEV of the consistent current
\bdi
\langle J^\mu_{\rm cons.}(x)\rangle \equiv \frac{i}{e}\frac{\delta W}{
\delta A_\mu (x)} =-\frac{e}{2\pi}\int d^2 y\frac{1}{\Box}(x-y)[
\partial^\mu \partial^\nu -\Box g^{\mu\nu}
\edi
\beq
-\frac{1}{2}(\epsilon^{\nu\alpha}\partial^\mu \partial_\alpha +
\epsilon^{\mu\alpha}\partial^\nu \partial_\alpha )A_\nu (y)
=:-\frac{e}{2\pi}\int d^2 yK^{\mu\nu}(x,y)A_\nu (y).
\eeq
The nonlocal kernel $K^{\mu\nu}(x,y)$ is Bose symmetric, $K^{\mu\nu}(x,y)
=K^{\nu\mu}(y,x)$, because it was derived as the functional derivative of
an effective action; but $\langle J^\mu_{\rm cons.}(x)\rangle$ is not
invariant under a gauge transformation $A_\mu \ra A_\mu +\partial_\mu
\lambda$. 

On the other hand, look at the expression
\beqa
\langle J^\mu_{\rm cov.}(x)\rangle &=& -\frac{e}{2\pi}\int d^2 y
\frac{1}{\Box}(x-y)[\partial^\mu \partial^\nu -\Box g^{\mu\nu} -
\epsilon^{\nu\alpha}\partial^\mu \partial_\alpha ]A_\nu (y) \no \\
&=:& \frac{-e}{2\pi}\int d^2 y \wt K^{\mu\nu}(x,y)A_\nu (y)
\eeqa
which differs from (3) by a local polynomial (see (8)).  Expression (4)  
is gauge invariant but not Bose symmetric (i.e. it may not
be obtained as the functional derivative of some effective action). 
The VEVs of 
the two currents lead to the consistent and
covariant anomalies
\beq
\partial_\mu \langle J^\mu_{\rm cons.}(x)\rangle = -\frac{e}{4\pi}
\epsilon^{\mu\nu}\partial_\mu A_\nu (x)=
\frac{e}{4\pi}
(\partial_0 A_1 (x) -\partial_1 A_0 (x))
\eeq
\beq
\partial_\mu \langle J^\mu_{\rm cov.}(x)\rangle = -\frac{e}{2\pi}
\epsilon^{\mu\nu}\partial_\mu A_\nu (x)
\eeq
which differ by a factor of 2 but are both given by the same gauge invariant 
expression. However, the addition of a local counterterm $\int d^2 xA_\mu (x)
C^{\mu\nu}A_\nu (x)$  to the 
effective action allows for a change of $\langle J^\mu_{\rm cons.}(x)
\rangle$ (where $C^{\mu\nu}=\left( \begin{array}{cc}a & b \\ b & c 
\end{array} \right)$),
\beqa 
\langle J^0_{\rm cons.}(x)\rangle &\ra & \langle J^0_{\rm cons.}(x)
\rangle  +aA_0 (x) +bA_1 (x) \no \\
\langle J^1_{\rm cons.}(x)\rangle &\ra & \langle J^1_{\rm cons.}(x)
\rangle  +bA_0 (x) +cA_1 (x)
\eeqa
where $a,b,c$ are arbitrary numbers. This allows for consistent anomaly
expressions like e.g. $\frac{-e}{2\pi} \partial_1 A_0 +c\partial_1 A_1$
(see (52)).
On the other hand, the covariant current and anomaly are fixed by
gauge invariance, as we shall see.

The consistent and covariant current VEVs (3) and (4) differ by the (local)
Bardeen-Zumino polynomial \cite{Bertl1,BZ1}
\beqa
P^\mu (x) &=& \langle J^\mu_{\rm cov.}(x)\rangle -\langle J^\mu_{\rm cons.}(x)
\rangle \no \\
&=& \frac{e}{4\pi}\int d^2 y\frac{1}{\Box}(x-y)(\epsilon^{\nu\alpha}
\partial^\mu \partial_\alpha -\epsilon^{\mu\alpha}\partial^\nu
\partial_\alpha )A_\nu (y) \no \\
&=& -\frac{e}{4\pi}\epsilon^{\mu\nu}A_\nu (x).
\eeqa
Other choices (7) for the consistent current VEV change $P^\mu$ by the local
and symmetric term $C^{\mu\nu}A_\nu$.

Althogether, we find that the consistent current has to obey Bose symmetry,
which is the abelian version of the Wess-Zumino consistency condition 
\cite{WZ1}, whereas the covariant current is determined by gauge invariance,
and there is no choice that obeys both Bose symmetry and gauge invariance.

Next we have to fix our conventions. We are in two-dimensional Minkowski
space-time $(x^0 ,x^1)\equiv x$ with the conventions
\beq
g^{\mu\nu} =\left( \begin{array}{cc}1 & 0 \\ 0 & -1 \end{array} \right)\, ,
\quad \epsilon_{01}=1 \, ,\quad
x^\pm =x^0 \pm x^1 .
\eeq
Our Lagrangian density is
\beq
{\cal L}=\bar\Psi (i\dsla -e\Asla P_+ )\Psi ,
\eeq
and we use the $\gamma$ matrix conventions
\beq
\gamma^{0} =\left( \begin{array}{cc}0 & 1 \\ 1 & 0 \end{array} \right)
\; ,\qquad
\gamma^{1} =\left( \begin{array}{cc}0 & -1 \\ 1 & 0 \end{array} \right)
\; ,\qquad
\gaf =\left( \begin{array}{cc}1 & 0 \\ 0 & -1 \end{array} \right)
\eeq
\beq
\gaf =\gamma^0 \gamma^1 \, , \quad P_\pm =\frac{1}{2}({\bf 1}\pm \gaf )
\eeq
and the currents ($J^\mu =\bar\Psi\gaM\Psi$)
\beqa
J^0 & =& (\Psi^\dagger_+ , \Psi^\dagger_- )\gamma^0 \gamma^0 
\left( \begin{array}{c} \Psi_+ \\ \Psi_- \end{array} \right) =
\Psi^\dagger_+ \Psi_+ +\Psi^\dagger_- \Psi_- \no \\
J^1 & =& (\Psi^\dagger_+ , \Psi^\dagger_- )\gamma^0 \gamma^1 
\left( \begin{array}{c} \Psi_+ \\ \Psi_- \end{array} \right) =
\Psi^\dagger_+ \Psi_+ -\Psi^\dagger_- \Psi_-
\eeqa
\bdi
J^\mu_+ =\bar\Psi \gaM P_+ \Psi =\frac{1}{2}(g^{\mu\nu}+\epsilon^{\mu\nu})
J_\nu
\edi
\beq
J^0_+ = J^1_+ =\Psi^\dagger_+ \Psi_+ =:J_+ .
\eeq
This leads to the Hamiltonian density
\beqa
{\cal H} &=& i\Psi^\dagger_+ \partial_0 \Psi_+ +i\Psi^\dagger_- \partial_0
\Psi_- -{\cal L} \no \\
&=& -i\Psi^\dagger_+ \partial_1 \Psi_+ +i\Psi^\dagger_- \partial_1
\Psi_- +eA_+ J_+ 
\eeqa
($\partial_\mu \equiv (\partial / \partial x^\mu)$), where
\beq
A_+ := A_0 +A_1 .
\eeq
Observe that in the Lagrangian and Hamiltonian no kinetic terms for the
gauge field occur, i.e., we shall treat $A_\mu$ as an external field throughout
the article.

Further, we shall frequently use the Baker-Campbell-Hausdorff (BCH) formula
for operators
\beq
e^A Be^{-A} =B+[A,B]+\frac{1}{2!}[A,[A,B]]+\ldots
\eeq

\section{What to expect}

It is a wellknown fact that both divergence anomaly and anomalous Gauss law 
commutator may be derived -- up to an overall constant -- by
cohomological methods via the so-called descent equations. For the
consistent anomaly this was done, e.g., in \cite{Sto1,Zu1}, 
and for the covariant
case in \cite{Tsu1}. Here we want to review these results briefly for
$d=1+1$ dimensions, because they will tell us what to expect in the
forthcoming computations.

All the two-dimensional anomalies and anomalous commutators may be
derived from the 3-dimensional Chern-Simons form
\beq
Q_3^0 (A,F)=\tr \, (AF -\frac{1}{3}A^3)
\eeq
where we deal with the general nonabelian case for the moment and the
trace is in color space. Here $A$ and $F=dA+A^2$ are differential forms on
coordinate space ($A=A_\mu dx^\mu$ etc.)

By substituting $A\ra A+v$, where $v$ is a one-form in group space, and by 
collecting powers in $v$
\beq
Q_3^0 (A+v, F)=Q_3^0 (A,F) +\tr \, vdA -\tr \, v^2 A -\frac{1}{3}\tr \, v^3 
=:\sum_k Q^k_{3-k}
\eeq
(where $k$ counts the degree in $v$), one obtains the expressions for the
consistent anomaly ($Q_2^1$) and Gauss law commutator ($Q_1^2$). However,
this result is not unique. The cohomological information is encoded in the 
relation
\beq
\delta_v Q^i_j =-dQ^{i+1}_{j-1}
\eeq
where $\delta_v$ is the exterior derivative on group space that generates
gauge transformations on $A,F$ and $v$ \cite{Bertl1}. As each $Q^i_j$ is only
fixed up to a total derivative, this, in turn, makes $Q^{i+1}_{j-1}$
ambiguous. In our case we have, e.g.,
\beqa
Q^1_2 =\tr \, vdA \quad &\ldots &\quad Q^2_1 =-\tr \, v^2 A \no \\
Q^1_2 =\tr \, dv A\quad &\ldots &\quad Q^2_1 =\tr \, vdv
\eeqa
where the two versions of $Q^1_2$ differ by a total derivative. In the
abelian case all higher powers of the one-forms $A,v$ vanish, and we find
for the divergence anomaly {\bf A} and Gauss law commutator $S$
\beq
{\rm\bf A}(x)\sim dA(x) \; , \quad S(x^1,y^1) \sim \delta ' (x^1 -y^1) \sim 0
\eeq
where $S$ is ambiguous and cohomologically equivalent to zero.

On the other hand, the covariant anomalies may be found by the expansion of
\beq
Q^0_3 (A+v, F+Dv)=Q^0_3 (A,F)+2\tr \, vF +\tr \, (vdv +v^2 A) 
-\frac{1}{3}\tr \, v^3 =:\sum_k \wt Q^k_{3-k}
\eeq
($D=d+[A,\; \; ]$), giving rise to the covariant anomalies
\beq
\wt {\rm\bf A}^a (x)\sim 2F^a (x) \; ,\quad \wt S^{ab}(x^1 ,y^1)
\sim D^{ab}(x^1) \delta(x^1 -y^1) 
\eeq
($a$ \ldots color index, $D^{ab}$ \ldots covariant derivative).
In the abelian case this simplifies to
\beq
\wt {\rm\bf A}(x) \sim 2dA \; ,\quad \wt S(x^1,y^1) \sim \delta ' (x^1 -y^1).
\eeq
Observe that, again, the covariant anomaly is twice the consistent one. 
Further, the $\wt Q^k_{3-k}$ are uniquely fixed by the requirement of gauge
covariance.

\section{Dirac vacuum, normal ordering and Schwinger terms}

Next we should briefly review the second quantization of the free theory.
In this section we closely follow the discussion of \cite{IST1}.
The free spinors obey the free Dirac equation
\beq
(\partial_0 \pm \partial_1 )\Psi_\pm =0
\eeq
and are therefore given by
\beq
\Psi_\pm (x^0, x^1)=\int_{-\infty}^\infty \frac{dk^1}{\sqrt{2\pi}}
b_\pm (k^1)e^{-ik^1 x^\mp}
\eeq
with dispersion $k^0 =\pm k^1$. Here the $b_\pm$ are the usual annihilation 
operators obeying the CAR
\beq
\{ b_+ (k^1),b^\dagger_+ (k'^1)\} =\{b_- (k^1),b^\dagger_- (k'^1)\} 
=\delta (k^1-k'^1)
\eeq
and all other anticommutators vanish. The free Hamiltonian reads
\beq
H_0 =\int dx^1 (-i\Psi^\dagger_+ \partial_1 \Psi_+ +i\Psi^\dagger_- \partial_1
\Psi_- )= \int dk^1 k^1 (b^\dagger_+ (k^1)b_+ (k^1)-b^\dagger_- (k^1)b_- (k^1))
\eeq
and is unbounded from below. This necessitates the introduction of the Dirac 
vacuum
\beqa
b_\pm (k^1) |\; \; \rangle_{\rm D} =0 \quad &\ldots &\quad \pm k^1 > 0 \no \\
b^\dagger_\pm (k^1) |\; \; \rangle_{\rm D} =0 \quad &\ldots &\quad \pm k^1 < 0
\eeqa
and the normal ordering w.r.t. the Dirac vacuum,
\beq
N b^\dagger_\pm (k^1)b_\pm (k'^1)=b^\dagger_\pm (k^1)b_\pm (k'^1)
\vert_{\pm k'^1>0}
-b_\pm (k'^1) b^\dagger_\pm (k^1)\vert_{\pm k'^1<0} .
\eeq
This normal ordering has the consequence of inducing the Schwinger term
in the commutator of normal-ordered currents (a fact that was already
known in the thirties \cite{Jor1}), 
\beq
NJ_+ (x^-)=\int\frac{dk^1 dk'^1}{2\pi}e^{-i(k'^1 -k^1)x^-}Nb^\dagger_+
(k^1)b_+ (k'^1)
\eeq
and its Fourier transforms
\beq
N\wt J_+ (p^1)=\int dx^- e^{ip^1 x^-}NJ_+ (x^-)=\int dk^1 Nb^\dagger_+
(k^1 -p^1)b_+ (k^1) .
\eeq
A straight forward computation reveals
\beq
[ N\wt J_+ (p^1),N\wt J_+ (q^1)]=\int_{-\infty}^\infty dk^1 (b^\dagger_+
(k^1 -q^1)b_+ (k^1 +p_1) - b^\dagger_+ (k^1 -p^1 -q^1)b_+ (k^1)) .
\eeq
At first sight it seems that by a shift of the integration variable this
expression can be made equal to zero. However, the operators in (34) act
on Fock states built out of the Dirac vacuum, and a shift of $k^1$ is therefore
in conflict with the definition of the Dirac vacuum (30) (i.e., the shift would
mix Dirac annihilation and creation operators).

One possibility for the further evaluation of (34) is to rewrite it as
its normal-ordered version plus some remainder. In the normal-ordered
expression the shift is legitimate, and the remainder precisely gives the
Schwinger term (see, e.g., \cite{ABH}-\cite{Vl1})
\beq
[N\wt J_+ (p^1),N \wt J_+ (q^1)] =p^1 \delta (p^1 +q^1) .
\eeq
Further possibilities for the evaluation of (34) are given e.g. in 
\cite{IST1,Man1} and lead to the same result. In coordinate space the 
Schwinger term reads
\beq
[ NJ_+ (x), NJ_+ (y)]=-\frac{i}{2\pi}\partial_{x^1}\delta (x^- -y^-).
\eeq
For $NJ_-$ an analogous result may be obtained (differing in sign from
(36)), but as $\Psi_-$ is 
noninteracting, it is unimportant in the sequel. 

For the interacting, positive chirality sector we shall identify the 
normal-ordered current $NJ_+$ with the consistent current (this
identification will be justified in Section 5). One immediate consequence 
is that $NJ_+$ cannot be gauge invariant (see Section 4).

Therefore, next we should find a candidate for the covariant current.
Precisely this was done in \cite{IST1} by introducing the concept of
kinetic normal ordering, what we want to review now.

The Dirac vacuum is introduced by splitting the fermionic Hilbert space into
eigenstates of the free Hamiltonian $-i\partial_1$ with positive and 
negative energy. Instead, one could split into eigenstates of the kinetic
momentum operator $-i\partial_1 +eA_1$ with positive and negative
kinetic energy. These eigenvalues (the kinetic energy) are gauge invariant,
measurable quantities, and the corresponding kinetic normal ordering
will indeed lead to gauge invariant operators (see Section 4). So let us
expand the free spinor $\Psi_+$ into annihilation operators of the free
and kinetic momentum:
\beq
\Psi_+ (x)=\int\frac{dk^1}{\sqrt{2\pi}}e^{ik^1x^1}b_+ (k^1, x^0)=
\int \frac{dk^1}{\sqrt{2\pi}}e^{ik^1x^1 -ie\lambda (x)}\wt b_+ (k^1, x^0)
\eeq
\beq
\lambda (x):= \int_{-\infty}^{x^1} d\bar x^1 A_1 (x^0 ,\bar x^1)
\eeq
$b_+(k^1 ,x^0)$ is the free time evolution of $b_+ (k^1)$, (27); 
$\exp (ik^1 x^1 -ie\lambda (x))$ is an eigenfunction of the kinetic
momentum $(-i\partial_1 +eA_1)$ with eigenvalue $k^1$. 

$b_+$ and $\wt b_+$ (and their Fourier transforms $\Psi_+$ and $\wt \Psi_+$)
are related by a gauge transformation,
\beqa
\wt \Psi_+ (x) &=& \int\frac{dk^1}{\sqrt{2\pi}}e^{ik^1x^1}\wt b_+ (k^1,x^0) =
e^{ie\lambda (x)}\Psi (x) \no \\
&\equiv & \Lambda^\dagger_+ (x^0)\Psi_+ (x)\Lambda_+ (x^0)
= \int\frac{dk^1}{\sqrt{2\pi}}e^{ik^1x^1}\Lambda^\dagger_+ (x^0)b_+ (k^1,x^0)
\Lambda_+ (x^0)
\eeqa
where $\Lambda_+ (x^0)$ implements the gauge transformation on Fock space and
is given by
\beq
\Lambda_+ (x^0)=e^{ie\int dx^1 \lambda (x)NJ_+ (x)}
\eeq
as may be shown easily by using the BCH formula (17) and the ETC relation
\beq
[NJ_+ (x^1_n),\ldots [NJ_+(x^1_1),\Psi_+ (y^1)]\ldots ] =(-1)^n  \Psi_+
(y^1)\prod_{i=1}^n \delta (x^1_i -y^1).
\eeq
Actually, $\wt \Psi [A_1 =0] =\Psi$, and therefore $\Lambda^\dagger_+ (x^0)$
implements the gauge transformation that goes to Couloumb gauge $A_1 =0$.

Now, following \cite{IST1}, we define kinetic normal ordering in complete
analogy with (31) as
\beqa
\wt N\, \wt b^\dagger_+ (k^1,x^0) \wt b_+ (k'^1,x^0) & = &
\wt b^\dagger_+ (k^1,x^0)
\wt b_+ (k'^1,x^0)\vert_{k'^1 >0}-\wt b_+ (k'^1,x^0)\wt b^\dagger_+ (k^1,x^0)
\vert_{k'^1 <0} \no \\
&=&
\Lambda^\dagger_+ (x^0)Nb^\dagger_+ (k^1,x^0)b_+ (k'^1,x^0)\Lambda_+ (x^0),
\eeqa
where the last equality follows at once. Using it we find for the kinetically 
normal ordered current and free Hamiltonian
\beq
\wt NJ_+ (x)=\Lambda^\dagger_+ (x^0)NJ_+ (x)\Lambda_+ (x^0)
\eeq
\beq
\wt N H_0 (x^0)=\Lambda^\dagger_+ (x^0)NH_0 (x^0)\Lambda_+ (x^0) -e\int dx^1
A_1 (x)\Lambda^\dagger_+ (x^0)NJ_+ (x)\Lambda_+ (x^0) .
\eeq
With the help of the BCH formula and the identity
\beq
[NH_0 (x^0),NJ_+ (x)] =-i\partial_0 NJ_+ (x)=i\partial_1 NJ_+ (x)
\eeq
(where we used the fact that $NJ_+ (x)$ is a Heisenberg operator of the 
free theory in the first step, and the conservation of the free current in the
second step) we finally get
\beq
\wt NJ_+ (x) =NJ_+ (x)+\frac{e}{2\pi}A_1(x)
\eeq
\beq
\wt NH_+ (x^0)=NH_+ (x^0)+\frac{e^2}{4\pi}\int dx^1 (A_1^2 (x)+2A_0 (x) 
A_1 (x))
\eeq
where $H_+$ is the Hamiltonian of the $\Psi_+$ field,
\beq
H_+ (x^0)=\int dx^1 (\Psi^\dagger_+ (x)(-i\partial_1)\Psi_+ (x) +
eA_+ (x)J_+ (x)).
\eeq
The reordering just adds local polynomials in the external gauge field and,
therefore, $NJ_+$ and $\wt NJ_+$ have the same Schwinger term. We shall identify
$\wt NJ_+$ with the covariant current in the forthcoming sections.

\section{Gauss law operator}

Before continuing, we want to emphasize again that the operators in the
last section were in the interaction picture of the full theory, and we
shall remain in the interaction picture in this section.

The Gauss law operator $G$ implements time independent gauge transformations
and may be found e.g. by requiring a covariant transformation of the time
independent Dirac equation,
\bdi
[\int dx^1 \lambda (x^0, x^1) G(x^0, x^1),(-i\partial_{y^1}+eA_1 (x^0,y^1))
\Psi_+ (x^0,y^1)]=
\edi
\beq
i\lambda (x^0,y^1)(-i\partial_{y^1}+eA_1 (x^0,y^1)
\Psi_+ (x^0,y^1).
\eeq
It reads
\beq
G(x)=\partial_1 \frac{\delta}{e\delta A_1 (x)}-iNJ_+ (x)\; , \quad
\wt G(x)=\partial_1 \frac{\delta}{e\delta A_1 (x)}-i\wt NJ_+ (x)
\eeq
where we defined the consistent ($G$) and covariant ($\wt G$) Gauss law
operators. Here $A_1 (x)$ is treated as a function of space only and the
time variable $x^0$ as a parameter, \\ 
i.e. $(\delta /\delta A_1 (x^0,x^1)) A_1 (x^0,y^1) = \delta (x^1 -y^1)$.

For the consistent Gauss law commutator we find from the Schwinger term (36)
(because $(\delta /\delta A_1 ) NJ_+=0$)
\beq
[G(x^0,x^1),G(x^0,y^1)]=\frac{i}{2\pi}\partial_{x^1}\delta (x^1 -y^1).
\eeq 
Further, we are able to reproduce the Fujikawa relation \cite{Fuji1} that
relates the commutator of Gauss law and Hamiltonian operators to the
consistent anomaly
\bdi
[G(x),NH_+(x^0)] = [\partial_{x^1}\frac{\delta}{e\delta A_1 (x)}-iNJ_+ (x),
NH_0 (x^0) +e\int dy^1 A_+ (x^0,y^1)NJ_+ (x^0,y^1)]=
\edi
\bdi
\partial_{x^1}\int dy^1 \delta (x^1 -y^1)NJ_+ (x^0,y^1) -i[NJ_+ (x), NH_0 (x^0)
] -
\edi
\beq
-ie\int dy^1 A_+ (x^0,y^1)[NJ_+ (x^0,x^1),NJ_+ (x^0,y^1)]=-\frac{e}{2\pi}
\partial_1 A_+ (x)
\eeq
where we used (45) and (36). This result is equal to the consistent anomaly 
(5) of the introduction up to a local (but Lorentz-noncovariant)
counterterm.

Next let us turn to the covariant Gauss law operator. First we observe that 
the covariant current is indeed gauge invariant (in contrast to the 
consistent one),
\bdi
[\wt G(x^0,x^1),\wt NJ_+ (x^0,y^1)] = 
\edi
\beq
[\partial_{x_1}\frac{\delta}{e\delta 
A_1 (x^0,x^1)}-iNJ_+ (x^0,x^1),NJ_+ (x^0,y^1) +\frac{e}{2\pi}A_1 (x^0,y^1)]
=0.
\eeq
For the covariant Gauss law commutator we obtain
\beq
[\wt G(x^0,x^1),\wt G(x^0,y^1)]=-\frac{i}{2\pi}\partial_{x^1}\delta(x^1 -y^1),
\eeq
i.e., it is minus the consistent Gauss law commutator (51).

In addition, we find that the covariantly regularized Hamiltonian $\wt NH_+$
is gauge invariant, too,
\bdi
[\wt G(x),\wt N H_+ (x^0)]=
\edi
\beq
[G(x^0,x^1),NH_+ (x^0)+\frac{e^2}{4\pi}
\int dy^1 (A_1^2 (x^0,y^1)+2A_0 (x^0,y^1)A_1 (x^0,y^1))]=0.
\eeq
Therefore, at least for the external field problem, the covariant anomaly 
cannot be inferred from a covariant version of the Fujikawa relation.

{\em Remark:} the covariant anomaly may be found from 
$[\wt G(x),\wt N H_+(x^0)]$,
when we treat $A_\mu$ as a dynamical field, i.e. include the gauge field
kinetic energy $H_{\rm g}=(-1/4)\int dx^1 F_{\mu\nu}F^{\mu\nu}=(1/2)\int
dx^1 E^2$, $E=\partial_0 A_1 -\partial_1 A_0$, into the Hamiltonian.
Using $[E(x^1),A_1 (y^1)]=-i\delta (x^1 -y^1)$ we find
\beq
[\wt G(x),H_{\rm g}(x^0)]=\frac{e}{2\pi}E(x)=-\frac{e}{2\pi}
\epsilon^{\mu\nu}\partial_\mu A_\mu (x),
\eeq
which is precisely the covariant anomaly (6). This shows that in ETCs 
the consistent and covariant anomalies have a somewhat different origin
(observe that $[G(x),NH_+ (x^0)]$ is not changed by the inclusion of
$H_{\rm g}(x^0)$, as $G(x)$ does not depend on $A_1$).
However, we shall continue to treat $A_\mu$ as an external, nondynamical 
field.  

\section{Time evolution and Heisenberg current operators}

In the sequel we shall assume that the gauge field $A_\mu (x)$ vanishes in the
remote past, $\lim_{x^0\to -\infty}A_\mu (x)=0$. The time evolution
operator is given by (see \cite{IST1})
\beqa
U(x^0,-\infty) &=& T\exp (-i\int_{-\infty}^{x^0}dx'^0 H_{\rm I}(x'^0)) \no \\
&=& \exp (-i\int_{-\infty}^{x^0}dx'^0 H_{\rm I}(x'^0) -iC(x^0))
\eeqa
where, in the consistent case,
\beq
H_{\rm I}(x^0)=e\int dx^1 A_+ (x)NJ_+ (x)
\eeq
\beqa
C(x^0) &=& \frac{1}{2}i\int_{-\infty}^{x^0}dy^0 \int_{-\infty}^{y^0}dz^0
[H_{\rm I}(z^0),H_{\rm I}(y^0)] \no \\
&=& -\frac{1}{4\pi}\int d^2 y d^2 z\theta (x^0 -y^0)\theta (x^0 -z^0)
\theta (y^0 -z_0)\delta (y^- -z^-)A_+ (y)\partial_{z^1}A_+ (z).
\eeqa
The perturbative expansion of the time-ordered exponential into ordinary
exponentials in (57) terminates at the quadratic order, because the
commutator of two interaction Hamiltonians $H_{\rm I}(x^0)$ is a c-number
for all times. Therefore, (57) is an exact result \cite{IST1}.

So let us compute the consistent current in the Heisenberg picture with the
help of the BCH formula (17)
\beqa
NJ_+^{\rm H} (x) &=& U^\dagger (x^0,-\infty)NJ_+ (x)U(x^0,-\infty) \no \\
&=& NJ_+ (x) + i\int_{-\infty}^{x^0} dy^0 [H_{\rm I} (y^0),NJ_+ (x)] \no \\
&=& NJ_+ (x) -\frac{e}{2\pi}\int d^2 y\theta (x^0 -y^0)\delta (x^- -y^-)
\partial_{y^1}A_+ (y).
\eeqa
First, let us prove that the current $NJ_+$ is indeed the consistent current.
Within perturbation theory, the VEV of the consistent current is defined as
the normalized functional derivative of the vacuum functional,
\beq
\langle J^\mu_{\rm cons.} (x)\rangle := \frac{i}{e}\frac{\delta}{\delta
A_\mu (x)}\ln Z[A_\mu]
\eeq
where
\beq
Z[A_\mu] =\langle 0,\, {\rm out}\, \vert \, {\rm in\,},0\rangle =
\langle 0,\, {\rm in}\, \vert U(\infty ,-\infty)\vert \, {\rm in\,},0\rangle .
\eeq
In our case, $\vert \, {\rm in}\, , 0\rangle $ is just the Dirac vacuum of the 
free theory, and $A_\mu (x)$ is now interpreted as a space-time function, 
$(\delta / \delta A_\mu (x))A_\nu (y)=\delta^\mu_\nu \delta^2 (x-y)$.
We find e.g. for $(\delta / \delta A_0 (x))$ (using again the BCH formula)
\bdi
\frac{1}{\langle 0,\, {\rm out}\, \vert \, {\rm in\,},0\rangle }\frac{i}{e}
\langle 0,\, {\rm in}\, \vert \frac{\delta}{\delta A_0 (x)}U(\infty,-\infty)
\vert \,{\rm in}\, ,0\rangle =
\edi
\bdi
\frac{1}{\langle 0,\, {\rm out}\, \vert \, {\rm in\,},0\rangle }\frac{i}{e}
\langle 0,\, {\rm in}\, \vert U(\infty,-\infty)\Bigl( \frac{\delta}{
\delta A_0 (x)}+[(i\int dy^0 H_{\rm I}(y^0) +iC(\infty)),\frac{\delta
}{\delta A_0 (x)}]
\edi
\bdi
+\frac{1}{2}[i\int dy^0 H_{\rm I} (y^0),[i\int dz^0 H_{\rm I} (z^0),
\frac{\delta}{\delta A_0 (x)}]]\Bigr) \vert \, {\rm in}\, ,0\rangle =
\edi
\bdi
\frac{1}{\langle 0,\, {\rm out}\, \vert \, {\rm in\,},0\rangle }
\langle 0,\, {\rm out}\, \vert NJ_+ (x) -\frac{e}{2\pi}\int d^2 y 
\theta(x^0 -y^0)\delta (x^- -y^-)\partial_{y^1}A_+ (y)\vert \, {\rm in}\, ,0
\rangle
\edi
\beq
\equiv \frac{1}{\langle 0,\, {\rm out}\, \vert \, {\rm in\,},0\rangle }
\langle 0,\, {\rm out}\, \vert NJ_+^{\rm H} (x)\vert \, {\rm in}\, ,0\rangle
\eeq
where
\beq
[i\int dy^0 H_{\rm I}(y^0),\frac{\delta}{\delta A_0 (x)}]=-iNJ_+ (x)
\eeq
\beq
[iC(\infty),\frac{\delta}{\delta A_0 (x)}]=\frac{ie}{4\pi}\int d^2 y
\epsilon (x^0 -y^0)\delta(x^- -y^-) \partial_{y^1}A_+ (y)
\eeq
\beq
\frac{1}{2}[i\int dy^0 H_{\rm I} (y^0),[i\int dz^0 H_{\rm I} (z^0),
\frac{\delta}{\delta A_0 (x)}]]=\frac{ie}{4\pi}\int d^2 y\delta (x^- -y^-)
\partial_{y^1}A_+ (y)
\eeq
\beq
\epsilon (x^0):=\theta (x^0)-\theta (-x^0)
\eeq
and we find a completely identical result for the other component
$(\delta /\delta A_1 (x))$ (remember that $J^0_+ =J^1_+ \equiv J_+$).
Actually, the derivation (63) remains the same for general in and out states,
and therefore the identification $J^\mu_{+{\rm cons.}}\equiv NJ_+^\mu$
holds for all $S$-matrix elements.

Next we want to compute the consistent anomaly
\beqa
\partial_\mu NJ^{\mu ,{\rm H}}_+ (x) &=& (\partial_0 +\partial_1 )(NJ_+ (x)-
\frac{e}{2\pi}\int d^2 y\theta (x^0 -y^0)\delta (x^- -y^-)\partial_{y^1}
A_+ (y)) \no \\
&=& -\frac{e}{2\pi}\partial_1 A_+ (x)=-\frac{e}{2\pi}(\partial_1 A_0 (x)
+\partial_1 A_1 (x)) 
\eeqa
where we used the fact that $\theta (x^0)\delta (x^-)$ is the (retarded)
Green function of the operator $(\partial_0 +\partial_1)$,
\beq
(\partial_{x^0} +\partial_{x^1} )\theta (x^0-y^0)\delta (x^- -y^-)=
\delta^2 (x-y).
\eeq
This result precisely coincides with the consistent anomaly (52) of Section 4.
 
Now we want to discuss the covariant current operator in an analogous manner
(here we just review the discussion of \cite{IST1}, where the covariant 
Heisenberg current and anomaly were already derived).

All the fermionic operators of Sections 3, 4 were Heisenberg operators of
the free Hamiltonian $NH_0$, therefore all the additional parts of $\wt NH$ 
must be treated as interaction terms,
\beq
\wt H_{\rm I} (x^0)=\wt NH(x^0)-NH_0 (x^0)=H_{\rm I}(x^0)+\frac{e^2}{4\pi}
\int dx^1 (A_1^2 (x)+2A_0 (x)A_1 (x))
\eeq
leading to the time evolution operator
\beq
\wt U(x^0,-\infty)=\exp (-i\int_{-\infty}^{x^0}dy^0 H_{\rm I} (y^0)-iC(x^0)
-iD(x^0))
\eeq
\beq
D(x^0)=\frac{e^2}{4\pi}\int d^2 y \theta (x^0-y^0)(A_1^2 (y)+2A_0 (y) A_1
(y)).
\eeq
For the adjoint action on fermionic operators the gauge field dependent phases
$iC(x^0), iD(x^0)$ are irrelevant, and we find for the covariant current in the
Heisenberg picture
\bdi
\wt NJ_+^{\rm H}(x)=\wt U^\dagger (x^0,-\infty)\wt NJ_+ (x)\wt U(x^0,-\infty)=
\edi
\beq
U^\dagger (x^0,-\infty)NJ_+ (x)U(x^0,-\infty)+\frac{e}{2\pi}A_1 (x)=
NJ_+^{\rm H} (x)+\frac{e}{2\pi}A_1 (x)
\eeq
and for the covariant anomaly
\beq
\partial_\mu \wt NJ_+^{\mu ,{\rm H}}(x)=-\frac{e}{2\pi}\partial_1 A_+ (x)
+\frac{e}{2\pi}(\partial_0 +\partial_1 )A_1 (x)=\frac{e}{2\pi}
(\partial_0 A_1 (x)-\partial_1 A_0 (x)).
\eeq
This is precisely the gauge and Lorentz invariant result (6) of the
introduction.

\section{Time evolution of the Gauss law operators}

In Section 4 the Gauss law operator was defined in (50), and there the gauge 
field was treated as a function of space only. For the time evolution we need 
a generalization to space-time functions. Following \cite{Pawl1,Pawl2} we
define
\beq
G(x)=\delta (x^1)-iNJ_+ (x) \; ,\quad \delta (x^1)=\int_{-\infty}^{\infty}dx'^0
\partial_{x^1}\frac{\delta}{e\delta A_1 (x'^0,x^1)}
\eeq
where $A_1(x)$ is now a space-time function, i.e. $(\delta /\delta A_1 (x))
A_1 (y)=\delta^2 (x-y)$. Obviously, $\delta(x^1)$ is just the generalization of
$\partial_1 (\delta /\delta A_1 (x^1))$ to space-time functions.

We are now in a position to compute the time evolution of the Gauss law 
operator, which we want to do for the consistent Gauss law operator (75)
first. For the time evolution of $\delta (x^1)$ we find
\bdi
U^\dagger (x^0,-\infty)\delta (x^1)U(x^0,-\infty) = \delta (x^1) + [(ie\int
d^2 y \theta (x^0 -y^0)A_+ (y)NJ_+ (y) +iC(x^0)), \delta (x^1)] 
\edi
\bdi
+\frac{(ie)^2}{2!}[\int d^2 y \theta (x^0 -y^0)A_+ (y)NJ_+ (y),[\int d^2 z
\theta (x^0 -z^0)A_+ (z)NJ_+ (z),\delta (x^1)]]=
\edi
\bdi
\delta(x^1) -i\int d^2 y \theta (x^0 -y^0 )\partial_{x^1}\delta (x^1 -y^1)
NJ_+ (y)+
\edi
\beq
\frac{ie}{2\pi}\int d^2 yd^2 z \theta (x^0 -y^0)\theta (x^0 -z^0)\theta
(z^0 -y^0)\delta(x^1 -z^1)\delta(y^- -z^-)\partial^2_{y^1}A_+ (y)
\eeq
where
\beq
[ie\int d^2 y \theta(x^0 -y^0)A_+ (y)NJ_+ (y),\delta (x^1)]=-i\int d^2 y
\theta (x^0 -y^0)\partial_{x^1}\delta (x^1 -y^1)NJ_+ (y)
\eeq
\bdi
[iC(x^0),\delta(x^1)]=
\edi
\beq
-\frac{ie}{4\pi}\int d^2 y d^2 z \theta (x^0 -y^0)
\theta (x^0 -z^0)\epsilon (y^0 -z^0)\delta(x^1 -z^1)\delta (y^- -z^-)
\partial^2_{y^1}A_+ (y)
\eeq
\bdi
\frac{(ie)^2}{2}[\int d^2 y \theta (x^0-y^0)A_+ (y)NJ_+ (y),[\int d^2 z
\theta(x^0 -z^0)A_+ (z)NJ_+ (z),\delta (x^1)]]=
\edi
\beq
\frac{ie}{4\pi}\int d^2 yd^2 z\theta (x^0-y^0)\theta (x^0-z^0)\delta
(x^1 -z^1)\delta (y^- -z^-)\partial^2_{y^1}A_+ (y).
\eeq
I.e., under time evolution the operator $\delta (x^1)$ acquires a contribution 
proportional to the fermionic current operator and a further contribution 
that is a nonlocal functional of the gauge field. This latter contribution 
itself consists of a trivial part, steming from the $[iC(x^0),\delta (x^1)]$
commutator, and a nontrivial part from the $[\int H_{\rm I},[\int H_{\rm I},
\delta (x^1)]]$ double commutator (what we mean by ``trivial'' and 
``nontrivial'' will become clear in the sequel). 

Now we want to investigate the time evolution of the commutator
\beq
[\delta (x^1),\delta (x'^1)]=0.
\eeq
We find ($U(x^0)\equiv U(x^0,-\infty)$)
\bdi
[U^\dagger (x^0)\delta (x^1)U(x^0),U^\dagger (x^0)\delta (x'^1)U(x^0)
]=
\edi
\bdi
(-i)^2\int d^2 yd^2 z\theta (x^0 -y^0)\theta (x^0 -z^0)\partial_{x^1}
\delta (x^1 -y^1)\partial_{x'^1}\delta (x'^1 -z^1)[NJ_+ (y),NJ_+ (z)]
\edi
\bdi
+\frac{ie}{2\pi}[\delta(x^1),\int d^2 yd^2 z\theta (x^0 -y^0)\theta
(x^0 -z^0)\theta (z^0 -y^0)\delta (x'^1 -z^1)\delta (y^- -z^-)
\partial^2_{y^1}A_+ (y)]
\edi
\bdi
-\frac{ie}{2\pi}[\delta(x'^1),\int d^2 yd^2 z\theta (x^0 -y^0)\theta
(x^0 -z^0)\theta (z^0 -y^0)\delta (x^1 -z^1)\delta (y^- -z^-)
\partial^2_{y^1}A_+ (y)]=
\edi
\beq
=\ldots \equiv 0,
\eeq
where the sum of the two commutators containing the $\delta (x^1)$,
$\delta (x'^1)$ precisely cancels the $[NJ_+,NJ_+]$ term.
Therefore, the commutator (80) remains unchanged under time
evolution! This happens, because the time evolution $U^\dagger \delta U$ 
of $\delta$
contains two nontrivial pieces that give nonzero contributions to the 
commutator (81), namely a piece containing the fermionic current $NJ_+$, (77),
and a gauge field piece that stems from the double commutator
$[\int H_{\rm I},[\int H_{\rm I},\delta (x^1)]]$, (79). These two
nontrivial contributions to the commutator precisely cancel each other
and make the commutator (81) vanish. There is another gauge field piece in
$U^\dagger \delta U$ steming from the $[C(x^0),\delta (x^1)]$ commutator, (78),
but this is trivial and gives no contribution to the $[U^\dagger \delta U,
U^\dagger \delta U]$ commutator, because
\beq
(\delta (x^1)\delta (x'^1) - \delta (x'^1)\delta (x^1))
C(x^0)\equiv 0.
\eeq
The time evolution of the consistent current was already derived in the 
last section, therefore we may now compute the time evolution of the anomalous
Gauss law commutator 
\bdi
[U^\dagger (x^0)(\delta (x^1)-iNJ_+ (x^0,x^1))U(x^0),
U^\dagger (x^0)(\delta (x'^1)-iNJ_+ (x^0,x'^1))U(x^0)]=
\edi
\bdi
(-i)^2 [NJ_+ (x^0,x^1),NJ_+ (x^0,x'^1)]  
\edi
\bdi
- i[U^\dagger (x^0)\delta (x^1)U(x^0),U^\dagger (x^0)NJ_+ (x^0,x'^1)
U(x^0)]\; -\; (x^1 \leftrightarrow x'^1) 
\edi
\beq
= \ldots =\frac{i}{2\pi}\partial_{x^1}\delta (x^1 -x'^1).
\eeq
Therefore, we find that the commutator of the Gauss law operator as well as
the commutators of all its components remain unchanged under time evolution
(see (51)).

Now let us briefly turn to the covariant Gauss law operator. The covariant 
time evolution operator $\wt U$ differs from the consistent one by the
phase factor $\exp (-iD(x^0))$, (72), and, therefore, $\wt U^\dagger (x^0)
\delta (x^1)\wt U(x^0)$ acquires an additional term
\beq
\wt U^\dagger (x^0)\delta (x^1)\wt U(x^0)=U^\dagger (x^0)\delta (x^1)U(x^0)
-\frac{ie}{2\pi}\int d^2 y\theta (x^0 -y^0)\delta (x^1 -y^1)
\partial_{y^1}A_+ (y).
\eeq
However, this additional term, steming from a phase factor, cannot change the
commutator. Therefore, we find that for the covariant Gauss law operator, 
too, the full commutator as well as the commutators of all components
remain invariant under time evolution.

At first sight, this result may seem surprizing. After all, it is a wellknown 
fact that the anomaly in anomalous gauge theories is related to a
nontrivial action of the functional derivative $(\delta /\delta A_1 (x))$
on the fermionic Fock space \cite{NS1}-\cite{Sem1}.

But we shall find that we precisely recover these features within our
approach. Let us look at the ``field strength'' operator $E$ itself,
\beq
E(x^1):= \int_{-\infty}^{\infty}dx'^0 \frac{\delta}{\delta A_1 (x'^0,x^1)}
\eeq
(instead of its derivative $\delta (x^1)=\partial_1 E(x^1)$ in the Gauss law).
Analogous to (76) it has the following consistent time evolution
\bdi
U^\dagger (x^0)E(x^1)U(x^0) =E(x^1)-ie\int d^2 y\theta (x^0 -y^0)\delta
(x^1 -y^1)NJ_+ (y)
\edi
\bdi
+\frac{ie^2}{4\pi}\int d^2 y d^2 z\theta(x^0 -y^0)\theta (x^0 -z^0)\epsilon
(z^0 -y^0)\delta (x^1 -z^1)\delta (y^- -z^-)\partial_{y^1}A_+ (y)
\edi
\beq
+\frac{ie^2}{4\pi}\int d^2 y d^2 z\theta(x^0 -y^0)\theta (x^0 -z^0)
\delta (x^1 -z^1)\delta (y^- -z^-)\partial_{y^1}A_+ (y)
\eeq
where the second line is from the $[iC(x^0),E(x^1)]$ commutator and the third
line is from the $[\int H_{\rm I},[\int H_{\rm I},E(x^1)]]$ double commutator.

The essential point now is that $E(x)$ has a nonvanishing VEV
\beq
\langle E(x)\rangle \equiv \langle 0,{\rm S}\vert E(x^1)\vert {\rm S},0\rangle
=\langle 0,{\rm IP}\vert E(x^1)\vert {\rm IP},0\rangle =
\langle 0,{\rm H}\vert U^\dagger (x^0)E(x^1)U(x^0)\vert {\rm H},0\rangle
\eeq
where the transformation from Schr\"odinger to interaction picture acts,
of course, trivially on $E(x^1)$. In our case the Heisenberg vacuum is just the
Dirac vacuum of the free theory, and therefore we get
\beq
\langle E(x)\rangle =\frac{ie^2}{2\pi}\int d^2 yd^2 z\theta (x^0 -y^0)
\theta (x^0 -z^0) \theta (z^0 -y^0)\delta (x^1 -z^1)\delta (y^- -z^-)
\partial_{y^1}A_+ (y).
\eeq
I.e., only the nonlocal gauge field part of the nontrivial time evolution
$U^\dagger EU$ occurs in the VEV, whereas the fermion current part has
zero VEV. Now the procedure in \cite{NS1} consists in defining a new
``field strength'' operator
\beq
\bar E(x):= E(x^1)+{\cal A} (x)\; ,\quad {\cal A}(x):=-\langle E(x)\rangle
\eeq
As $E(x^1)$ is just an ordinary functional derivative, $\bar E(x)$ may be
interpreted as a covariant functional derivative that is invariant
under $U(1)$ phase transformations of the time evolution operator,
$U(x^0)\ra \exp (if[A](x^0))U(x^0)$, where $f[A](x^0)$ is an arbitrary
functional of $A_\mu$. ${\cal A}$ is the corresponding $U(1)$-connection
and gives rise to the curvature
\beq
{\cal F}(x^1,y^1):=[\bar E(x^0,x^1),\bar E(x^0,y^1)]=-\frac{ie^2}{4\pi}
\epsilon (x^1 -y^1).
\eeq
${\cal F}(x^1,y^1)$ is solely determined by the double commutator 
contribution to (86) (the third line), because the other part stems from a 
pure phase $iC(x^0)$. Generally ${\cal A}(x)=\int d^2 y{\cal A}(x,y)
A_+ (y)$, and only the antisymmetric part of the kernel ${\cal A}(x,y)$
determines ${\cal F}(x^1,y^1)$, whereas the symmetric part is a pure
(functional $U(1)$) gauge. 

When we use this new, covariant functional derivative $\bar E$ in the Gauss
law operator,
\beq
\delta (x^1)\ra \bar \delta (x)=\frac{1}{e}\partial_1 \bar E(x),
\eeq
then the consistent commutator anomaly is just doubled,
\beq
[\bar G(x^0,x^1),\bar G(x^0,y^1)]=\frac{i}{\pi}\partial_{x^1}\delta (x^1 -y^1)
\; ,\quad \bar G (x):= \bar \delta (x)-ieNJ_+ (x).
\eeq
On the other hand, the covariant commutator anomaly vanishes,
\beq
[\bar{\wt{G}}(x^0,x^1),\bar{\wt{G}}(x^0,y^1)]=0 \; ,\quad \bar{\wt{G}}(x):=
\bar \delta (x) -ie \wt NJ_+ (x).
\eeq
This means that the Gauss law operator $\bar{\wt{G}}(x)$ itself is gauge 
invariant. Completely analogous results were found in \cite{NS1}.

{\em Remark:} As $[\bar{\wt{G}},\bar{\wt{G}}]=0$, $[\bar{\wt{G}},\wt NH]=0$,
the full theory (including the gauge field) may be quantized. This was
done in \cite{NS1}, and the resulting quantum field theory was found to break
Lorentz invariance even at a physical level. However, this discussion is
beyond the scope of our article, where the gauge field is treated as an
external field throughout.

The main result of this investigation that we want to emphasize again is
the fact that the nontrivial VEV of the ``field strength'' $E$, (88), and
the resulting functional curvature (90) are perfectly compatible with
the canonical time evolution of the ``field strength'' commutator
\bdi
[U^\dagger (x^0)E(x^1)U(x^0),U^\dagger (x^0)E(y^1)U(x^0)]=
\edi
\beq
U^\dagger (x^0)[E(x^1),E(y^1)]U(x^0)=0.
\eeq 

\section{Summary}

We have discussed all the anomalous structure of chiral QED$_2$ by applying
the formalism of canonical second quantization to the fermion field.
The introduction of the Dirac vacuum and normal ordering, and the
resulting Schwinger term in the current-current commutator were the
essential steps in this procedure, and the consistent and covariant
divergence and commutator anomalies are just consequences of these
fundamental concepts. By splitting the fermionic Hilbert space into
positive and negative energy sectors w.r.t. the free and kinetic momentum,
respectively, we could identify the consistent and covariant currents and 
anomalies (where we used the results of \cite{IST1} for the kinetic
normal ordering and covariant current).

Further, we computed the consistent and covariant Gauss law operators and 
the commutators of all its components both in the interaction and 
Heisenberg pictures. We found that the time evolution of the commutators
is canonical. Especially the ``field strength'' commutator $[E(x^1),
E(y^1)]=0$ remains zero under time evolution. This is compatible
with a nontrivial VEV $\langle E(x)\rangle \ne 0$, because the time evolution
of $E(x^1)$ contains two nontrivial pieces (a gauge field piece and a fermion 
current piece) that cancel each other in the commutator.

In addition, we found that the consistent and covariant time evolution
of the ``field strength'' $E(x^1)$ and of the gauge field part of the Gauss
law, $\partial_1 E(x^1)$, only differ by a trivial term (a phase in the
time evolution operator). Therefore, we obtained the same results for their
consistent and covariant commutators.

Here, of course, the question arises what can be learned from our computations
for more difficult chiral gauge theories. For chiral QCD$_2$ it remains 
true that normal ordering is sufficient to render all operators and VEVs
finite. E.g., the time evolution operator may be computed analogously
to (57). The perturbation series does not terminate for chiral QCD$_2$,
but only the first few terms are relevant for anomalies. Therefore, an
analogous discussion should be possible for chiral QCD$_2$, and it should
lead to analogous results (see \cite{IST2} for the covariant current 
and anomaly).

On the other hand, for $d=4$ the situation is more involved. There, even after
normal ordering some operator products remain singular and need 
regularization. This regularization has to be performed e.g. for the
time evolution operator and prevents a direct applicability of our simple 
computations and conclusions. 

%
%

\section*{Acknowledgement}

The author thanks R. Jackiw for the opportunity to join the Center of 
Theoretical Physics at MIT, where this work was performed, and the
CTP members for their hospitality.
Further thanks are due to R. Bertlmann, R. Jackiw and J. Pawlowski 
for helpful discussions.

This work is supported by a Schr\"odinger stipendium of the Austrian FWF.

\end{document}